\documentclass[twocolumn,showpacs,preprintnumbers,amsmath,amssymb,floatfix]{revtex4}

\usepackage{graphicx}
\usepackage{epstopdf}
\usepackage{dcolumn}
\usepackage{bm}

\usepackage[colorlinks=true, pdfstartview=FitV, linkcolor=blue, citecolor=blue, urlcolor=blue]{hyperref}

\begin{document}

\title{Compression and stretching of a self-avoiding chain in cylindrical nanopores}

\author{Suckjoon Jun}
 \affiliation{FAS Center for Systems Biology, Harvard University, 7 Divinity Ave., Cambridge, MA 02138}
 \author{D. Thirumalai}
 \affiliation{Biophysics Program, IPST and Department of Chemistry and Biochemistry, University of Maryland at College Park, College Park, Maryland 20742}
 \author{Bae-Yeun Ha}
  \affiliation{Department of Physics and Astronomy, University of Waterloo, Waterloo, Ontario N2L 3G1, Canada}

\date{\today}

\begin{abstract}
Force-induced deformations of a self-avoiding chain confined inside a cylindrical cavity, with diameter $D$, are probed using molecular dynamics  simulations, scaling analysis, and analytical calculations.  We obtain and confirm a simple scaling relation $-f \cdot D \sim R^{-9/4}$ in the strong-compression regime, while for weak deformations we find $f \cdot D = -A(R/R_0) + B (R/R_0)^{-2}$ , where $A$ and $B$ are constants, $f$ the external force, and $R$ the chain extension (with $R_0$ its unperturbed value).  For a strong stretch, we present a universal, analytical force-extension relation. Our results can be used  to analyze the behavior of biomolecules in confinement.
\end{abstract}

\maketitle

Technological advances have made it possible to visualize and manipulate individual polymer molecules that are trapped in nanopores~\cite{Austin04, Austin05, Reccius05, Dekker07}. This enables one to test theoretical predictions on confined polymers,  shedding new insights into their static and dynamic properties. Besides their technological importance, confined polymers are relevant in a number of biological processes. In the context of proteins, a newly synthesized protein exits the ribosome through a narrow cylindrical pore~\cite{Venki01} assisted, perhaps, by a tensile force~\cite{Klimov02}. Both chaperone-assisted protein folding and protein degradation by proteosome involve encapsulation of proteins in cylindrical pores, in which they experience compression and stretching forces~\cite{Martin08,Thirumalai01}.  Furthermore, many bacterial species are rod-shaped and even filamentous, and their chromosomes are highly compressed inside the cell and yet well-organized~\cite{Stavans06,Jun06}. Thus, the study of cylindrically confined polymers is not only of practical importance but also is a first step towards understanding several fundamental biological processes.

Surprisingly, compared to the progress on the experimental side, and despite its fundamental and practical importance, theoretical understanding of force-induced chain deformations in a pore remains behind. In the literature, analysis of force-compresion-extension measurements heavily relies on and is often limited to simple scaling arguments~\cite{Austin04, Austin05, Reccius05}.
  
The main purpose of this Letter is to present a  quantitative picture of the interplay between confinement and force-induced chain deformations (from strong compression to almost full stretching). To this end, we combine a scaling analysis, molecular dynamics (MD) simulations, and a systematic theoretical approach, and obtain force-compression-extension relations in almost the full range of chain deformations. 
Our results thus provide a quantitative basis for experiments involving  biomolecules (e.g., dsDNA and chromosomes) in strong confinement as in nano- and micro-channels.

\begin{figure}[tp]
  \centering
  \includegraphics[width=8.0cm]{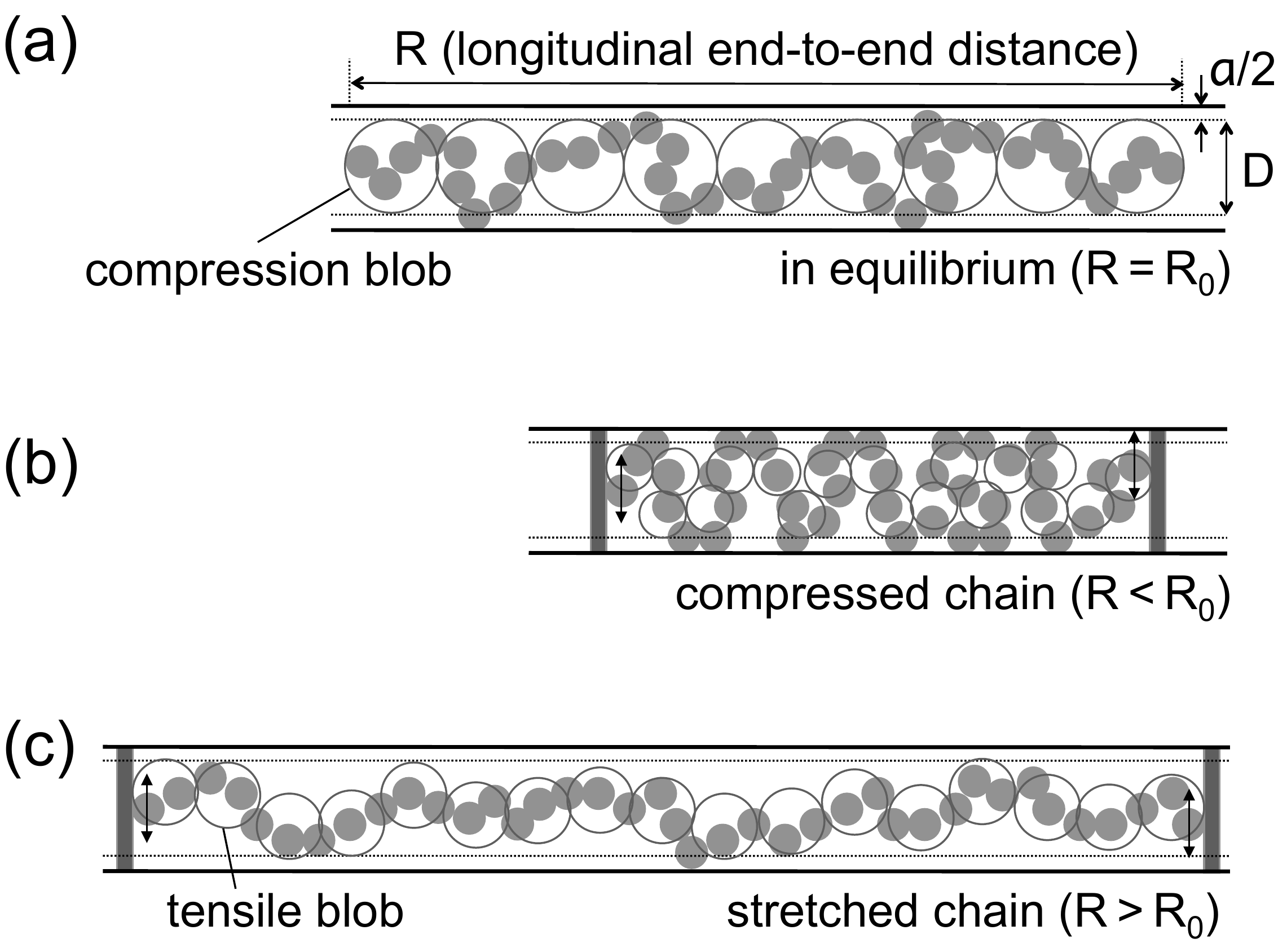}
    \caption{Self-avoiding polymer in a nanopore (a).  Simulation setting for compression (b) and stretching (c); the two chain ends are confined to, but can freely move within, the confining ``piston" walls.  By measuring the pressure exerted on the piston walls, we obtain force-compression-stretching relations, for a given chain size $N$ and a pore width $D$.}
   \label{fig:setup}
\end{figure}

Consider a self-avoiding polymer with $N$ monomers in a cylindrical pore with diameter $D$, as illustrated in Fig.~\ref{fig:setup}, with $R$ being the longitudinal end-to-end distance the chain.  Owing to the simultaneous presence of self-avoidence and confinement, the chain resembles a linear string of ``compression blobs"~\cite{Brochard77,deGennesBook}; inside each blob, the effect of confinement is not significant.  Because of the clear separation between two length scales ($>D$ and $< D$), the conventional Flory approach may fail to capture the correct $D$ dependence of relevant quantities (see below). The essential features of the blob-scaling approaches~\cite{Brochard77,deGennesBook} can be reproduced by the following ``renormalized'' free energy~\cite{endnote_Flory},
\begin{equation}
\label{eq:freeBY}
\beta\mathcal{F}(R,D) = A \frac{R^2}{(N/g)D^2}+ B \frac{D(N/g)^2}{R} ,
\end{equation}
where $g$ is the number of monomers inside a compression blob of diameter $D$, {\i.e.}, $g \simeq (D/a)^{5/3}$, $A$ and $B$ are constants, and $a$ is the monomer size.   (Throughout this paper, $\beta =1/k_BT$ with $k_BT$ being the thermal energy; unless otherwise indicated, $a$ is the unit of length.) The first term in Eq.~\ref{eq:freeBY} describes connectivity of blobs, while the second term, which can be rewritten as $\sim (N/g) \times$ blob density, represents the mutual repulsion between the blobs, thus ensuring linear ordering.  The second term is equivalent to correctly assigning $k_BT$ to a binary collision of two blobs, independent of their size~\cite{deGennesBook,Jun07}.      

The free energy in Eq.~\ref{eq:freeBY} produces \emph{not only} the expected equilibrium chain size $R_{ 0} \sim N D^{-2/3}$~\cite{endnote_uncertainty}, when $\mathcal{F}$ is minimized, \emph{but also} the correct confinement free energy 
$\beta \mathcal{F}_\mathrm{conf} = \beta \mathcal{F} (R=R_0,D) \sim N/g \sim N D^{- 5/3}$, which is linear in $N$~\cite{deGennesBook,Brochard77}.   The free energy cost for linear ordering is correctly counted as $\sim k_BT$ per blob ($D$- independent).  Also, the resulting effective Hookean spring constant of the chain, $k_\mathrm{eff} = \left({\partial^2 \mathcal{F}/\partial R^2} \right)_{R_0}\sim N^{-1}D^{-1/3}$ is consistent with the previous scaling result~\cite{Brochard77, Austin04}. Note that, without the renormalization trick in Eq.~\ref{eq:freeBY}, Flory theory (as employed in~\cite{Reccius05}; see also~\cite{endnote_Flory}) will lead to a $D$-independent (thus qualitatively inaccurate) spring constant $k_\mathrm{eff}$.

An important consequence of  Eq.~\ref{eq:freeBY} is that it predicts a universal scaling relation for (external) force-compression-extension ($R$) relation,
\begin{equation}
\label{eq:force}
D \beta f
= D  \frac{\partial }{\partial R} \left( \beta \mathcal{F} \right)= 2A \left(\frac{R}{R_{ 0}}\right) - B\left(\frac{R}{R_{ 0}}\right)^{-2}
.\end{equation}

To test the validity of Eq.~\ref{eq:force} (thus Eq.~\ref{eq:freeBY} as well), we have carried out molecular dynamics simulations using ESPResSo~\cite{espresso04a} for a wide range of parameters: $D = 4, 5, \ldots, 15$ and $N=128, 256, 512$ (see Fig.~\ref{fig:setup}).  In ESPResSo, monomers (or beads) are connected by finite extensible nonlinear elastic (FENE) (spring) bonds with a Weeks-Chandler-Andersen (WCA) potential for excluded-volume interactions (monomer-monomer as well as monomer-wall). We confined the two end beads of the chain by  ``piston'' walls at the cylinder ends (Fig.~\ref{fig:setup}(b)\&(c)), while allowing the ends to move freely in the plane of the piston wall.  By varying the wall-to-wall distance (thus $R$), we obtained the longitudinal force $f(R)$ to keep $R$ fixed.

\begin{figure}[tp]
  \centering
  \includegraphics[width=8.0cm]
  {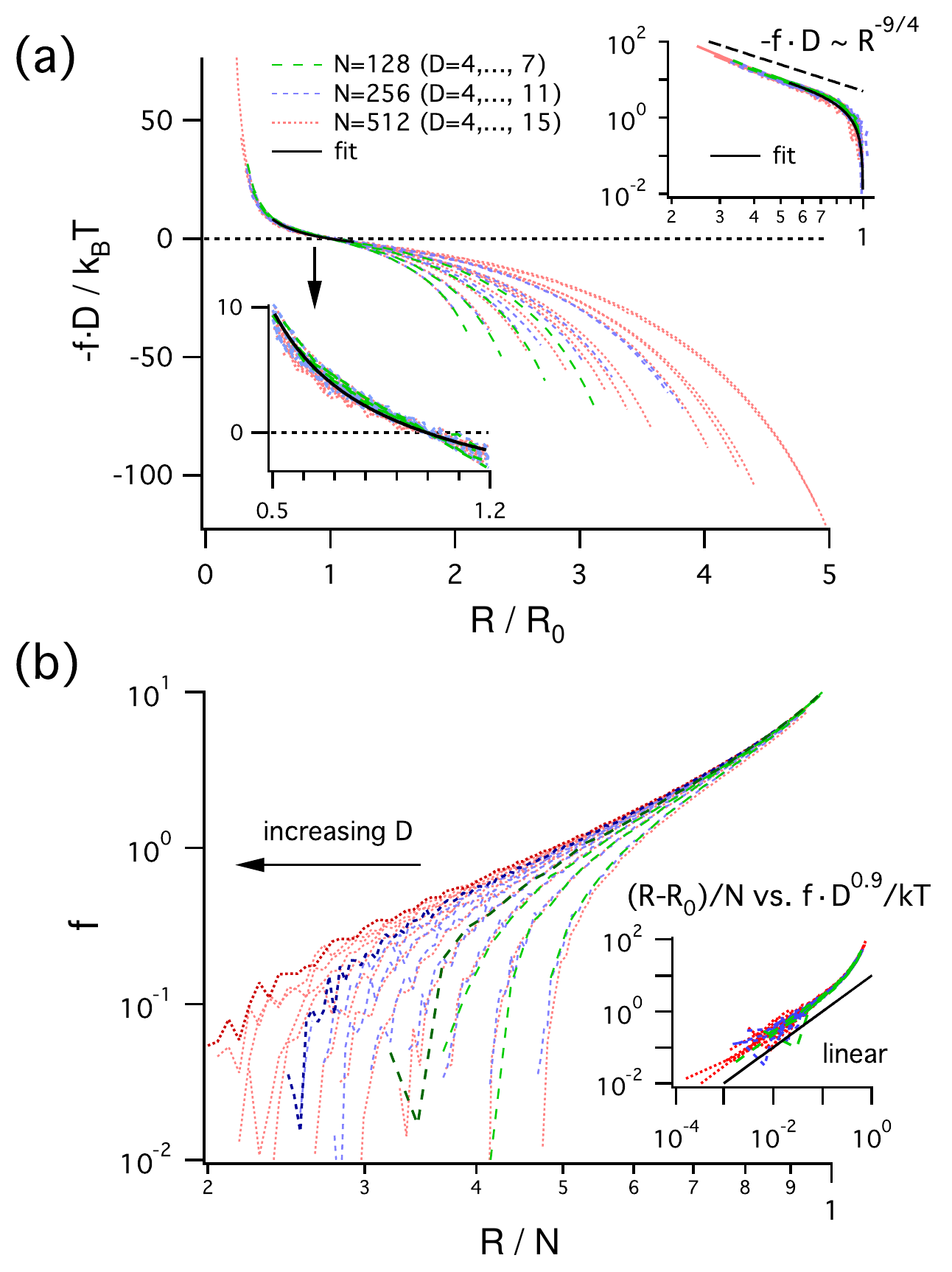}
  \caption{[Color online; (a) and (b) have the same color scheme] Simulation (dotted lines) and scaling (solid lines) of force-compression-stretching of a chain in a cylinder. (a) $R/R_0$ vs. $f \cdot D/k_BT$.   All 24 simulation curves collapse onto a single universal curve for $R < 1.2 R_0$ [$N=128 \ (D=4, 5, 6, 7), N=256 \ (D=4, 5, \ldots, 11), N=512 \ (D=4, 5, \ldots, 15)$].  The solid line is the scaling result using Eq.~\ref{eq:force} in the range of $0.5 < R /R_{ 0} < 1.2$; by a global fit to the simulation, we obtained $2A=2.84 \pm 0.03$ and $B = 2.81 \pm 0.01$ (\i.e., $2A \approx B$). The lower-left inset shows the fitted region. The upper-right inset is the log-log plot for $R < R_{ 0}$.  (b) $R /N$ vs. $|f|$ for force-stretching ($R > R_{ 0}$). The inset shows the linear Hookean regime for weak force (see text).  The sudden departure from the linear relation as $R /N \rightarrow 1$ reflects the non-universal aspect of the bond potentials in simulations.}
  \label{fig:sim_force}
\end{figure}

Figure~\ref{fig:sim_force}(a) summarizes our simulation results, where all 24 curves have been rescaled according to Eq.~\ref{eq:force}. The data perfectly collapse onto a single master curve (the solid line) for $R < R_{ 0}$. 
Moreover, in the range of $0.5 < R /R_{0} < 1.2$, Eq.~\ref{eq:force} fits the rescaled data very well, thus confirming the scaling prediction of Eq.~\ref{eq:freeBY} [see the lower-left inset in Fig.~\ref{fig:sim_force}(a)]. However, as the compression becomes strong ($R /R_{0} \ll 1$), the blobs break into smaller ones [see Fig.~\ref{fig:setup}(b)], and the correct form of free energy is $\beta \mathcal{F} \sim (R_g^3/V)^{1/(3\nu-1)}$, where $R \sim N^\nu$ is the radius of gyration of a corresponding unconfined chain (with $\nu \approx 3/5$ the Flory exponent) and $V \sim D^2 R$ the confining volume~\cite{Jun07}. From this, we obtain 
\begin{equation}
\label{eq:semidilute}
-f \cdot D \sim \left(\frac{R_0}{R}\right)^\frac{3\nu}{3\nu-1} \sim \left(\frac{R_0}{R}\right)^\frac{9}{4}
\end{equation}
[see the upper-right inset in Fig.~\ref{fig:sim_force}(a)].

\begin{figure}[tp]
  \centering
  \includegraphics[width=8.0cm]{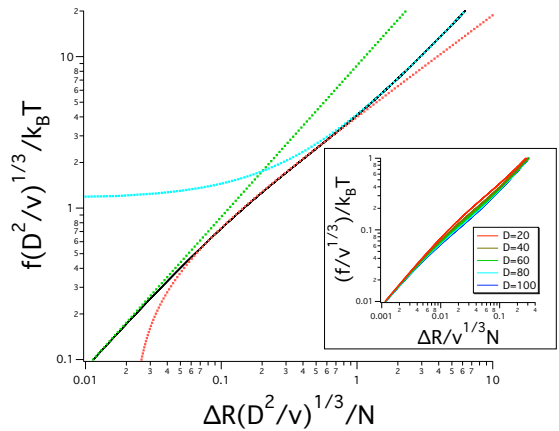}
  \caption{(Color online)  Rescaled force-extension relation in the limit $N \rightarrow \infty$.
  Three distinct regimes are identified and presented by the dotted lines (green, red, and cyan), which are  linear, power-law, and linear fits to the force-extension relation in their respective regimes.  The power-law analysis leads to $k_\text{eff} \sim 1/D^{1/3}N$.   }
  \label{fig:scaled-force-extension}
\end{figure}

Importantly, Fig.~\ref{fig:sim_force} shows the limitations  of Eq.~\ref{eq:force}; when the chain is stretched appreciably from its equilibrium length ($R \agt 1.2 R_0$), the force-extension curves (FECs) do not collapse onto a single universal plot.  This is not unexpected since an additional length scale, namely the tensile blob size $\xi$ ($<D$)~\cite{deGennesBook}, is relevant at high $f$, as indicated in Fig.~\ref{fig:setup}(c). This also signals the onset of a $D$-independent regime at large $f$, as the interactions between the chain and the wall are negligible [see Fig.~\ref{fig:sim_force}(b) and Fig.~\ref{fig:scaled-force-extension}].

In Fig.~\ref{fig:sim_force}(b), we show FECs  for $R > R_0$ by rescaling  $R$ by $N$. Since $R/N$ denotes the extension per chain segment, for the same pore diameter $D$, the force extension curves tend to collapse
onto one another, as expected.  In a recent simulation study~\cite{Arnold07b}, the effective spring constant $k_\mathrm{eff}$ was shown to vary from $k_\mathrm{eff} \sim 1/N D^{0.7}$ to $1/N D^{1.3}$ for the intermediate chain length (up to $N=2000$).  This differs from the scaling result $k_\mathrm{eff} \sim 1/N D^{1/3}$ ~\cite{Brochard77}, which could be reached only when  $N \sim 10^4$~\cite{Arnold07b}.  Our results are also in good agreement with Arnold \emph{et al.} and we find the linear Hookean collapse for $k_\mathrm{eff} \sim 1/ND^{0.9 \pm 0.2}$ [$f \sim k_\mathrm{eff} (R-R_0)$; inset of Fig.~\ref{fig:sim_force}(b)].  The data cannot be fit using the large $N$ result, $k_\mathrm{eff} \sim 1/ND^{1/3}$~\cite{endnote_dePablo}.

To further elucidate the subtleties of the FECs in the presence of confinement, we have also performed analytical, self-consistent calculations following the theoretical methods described in~\cite{Thirum07,Thirum05}.  Our motivation is to derive the FEC relation in the large-$N$ limit, which can be reached experimentally (if not computationally; see above).
In the continuum limit, the distribution of monomers at $\mathbf{r}(s)$ can be represented by the following Hamiltonian
\begin{eqnarray} 
\beta {\cal H} &=& {3 \over 2 a} \int_0^L\!\dot {\bf r}^2(s) d s  + {v \over 2} \int_0^L\!\int_0^L  ds ds' \delta ({\bf r}(s')-{\bf r}(s)) \nonumber \\
&& \quad-\beta f\int_0^L \dot z(s) ds
,\end{eqnarray}
where $s$ is the contour length ($0 \le s\le L=Na$),  $\dot {[...]}= \partial [...]/\partial s$, and $v$ the excluded volume parameter.  Confinement effects will be taken into account through the boundary condition  that the probability of finding any monomer on the wall is zero.  

\begin{widetext}
We replace $\mathcal{H}$ by the simpler reference $\mathcal{H}_1$ given by 
 $\beta {\cal H}_1 = {3 \over 2 a_1} \int_0^L \dot {\bf r}^2(s) d s - \beta f \int_0^L \dot z(s) ds$
and choose an optimal value of $a_1$ -- instead of minimizing the free energy as in the Flory approach --
so that $R$ obtained using $\mathcal{H}_1$, \i.e., $R \equiv \left< z_L -z_0 \right> = \beta La_1 f/3$, coincides with $R$ calculated using $\mathcal{H}$ to first order in $v$~\cite{Thirum07,Thirum05}.
Additionally, we use the ``ground-state dominance" approximation, which is valid in the large $N$ limit~\cite{deGennesBook}.   The optimal $a_1$ satisfies  the self-consistent equation (SCE):
\begin{eqnarray}
\label{SCE1}
La_1^{3/2} \left(\frac{1}{a}-\frac{1}{a_1} \right) &=& \sqrt{\frac{6}{ \pi}} \frac{1.76}{\pi} \frac{v}{\alpha_0^2 D^2} \int_0^L ds (L-s) \sqrt{s} \exp \left[ -\frac{1}{6} sa_1 (\beta f)^2\right] \nonumber \\
&=&
0.4 {v\over D^2} \left[ {18 e^{-\beta^2 f^2La_1} \over L^{1/2} \beta^4 f^4 a_1^2 } + { \sqrt{6 \pi}(-9 + L \beta^2 f^2 a_1) \ \mathrm{erf}  (f \sqrt{La_1/6} )\over  \beta^5 f^5 a_1^{5/2}}\right]
,\end{eqnarray} 
\end{widetext}         
where $\alpha_0 \approx 2.40$ is the smallest (first) zero of the zeroth-order Bessel function of the first kind $J_0(x)$, and $\mathrm{erf} (x)$ is the error function.  When combined with  $R = \beta La_1f/3$, the solution of the SCE determines the equilibrium chain length along the pore axis.

For $f = 0$, the SCE leads to the equilibrium chain size $R_0 = \sqrt{\left< (z_L-z_0 )^2 \right>}_{f=0} \sim L (\frac{va}{D^2})^{1/3}$, which is consistent with the previous result~\cite{Brochard77,Thirum05}.   For $f >0$, the SCE can be solved numerically.  A few comments are in order:  First, the excluded volume parameter $v$ in the SCE can be considered as a fitting parameter (see Ref.~\cite{Thirum07}). The value of $v$ can be adjusted to ensure the best global fit to simulations or experimental data. Second, our SCE does not reflect the kinetic aspects of force-extensions; chain reorientation, allowed in our SCE, is kinetically suppressed for strong confinement (It is strictly forbidden in our simulations.)  Finally, because of the ground-state-dominance approximation, the SCE becomes more accurate for large $L$.  In the limit $L\rightarrow \infty$, chain reorientation becomes irrelevant in that chain alignment in the force-direction is ensured for $f \agt k_BT/R_{ 0}$ ($\rightarrow 0$ as $L \rightarrow \infty$).  In this case, $v$ dependence enters the FEC only via the combination $D^2v$ and, thus, can be absorbed into an effective $D$ (see below). Our focus below is thus the large $L$ behavior of our SCE. 

In the large $L$ limit, the SCE in Eq.~\ref{SCE1} reduces to  $a_1^3 \left( {1\over a}-{1\over a_1}\right) \approx 0.44 {v \over D^2} {1\over (\beta f)^3}$.  
This equation implies that $a_1$ is a function of $f D^{2/3}/v^{1/3}$: $a_1 =a_1 (f D^{2/3}/v^{1/3})$.  As a result, $\left(\Delta R/N \right) D^{2/3}/v^{1/3}= \left( \beta a_1 f /3 \right)D^{2/3}/v^{1/3}$ is a function of $f D^{2/3}/v^{1/3}$, where $\Delta R=R-R_0$.  Thus, the FECs collapse onto each other in a $fD^{2/3} /v^{1/3}$-$\left(\Delta R /N\right)  D^{2/3}/v^{1/3}$ plot. 
The analytic results justify the scaling plot in Fig.~\ref{fig:sim_force}, where the renormalized $D_R \sim D v^{1/2}$.  In Fig.~\ref{fig:scaled-force-extension}, we plot $\Delta R D^{2/3}/v^{1/3}$ as a function of $f D^{2/3}/v^{1/3}$ in the limit $N \rightarrow \infty$.  All the curves for different values of $D$ collapse onto each other (as found in the simulations) for the reason described above. (Also see the inset in  Fig.~\ref{fig:scaled-force-extension} obtained without $D$ rescaling. This suggests that a more strongly confined chain behaves as a stronger spring.)   

Three different force-extension regimes are identified in Fig.~\ref{fig:scaled-force-extension}: linear (green dotted), power-law (red) and linear (cyan). The good linear fit to the very narrow linear regime indicates a weak $D$-dependence of the force-extension relation~\cite{endnote_I}. For the power-law fit in the second regime, we have chosen $R D^{2/3} \sim \text{const.} + \left(f D^{2/3}\right)^\alpha$ with $\alpha \approx 1.5$.  From this, $k_\text{eff}$ has been estimated to vary as $D^{-2(1-\alpha)/3} \approx D^{-1/3}$.  The seemingly-perfect fit to the force-extension in this regime justifies the scaling result  $k_\text{eff} \sim 1/D^{1/3}N$.  However, the force-extension in this regime is nonlinear, in contrast to what one may expect from the scaling approach~\cite{Brochard77}.  Only over a narrow parameter range can the linearity be recovered. We note, however, that this power-law regime represents a much wider parameter space than the initial linear regime. The large-$f$ linear regime is where the force-extension becomes $D$-independent is due to almost full stretching of chain by large force.

Our results suggest that relaxation dynamics of a chain confined in a narrow channel will  show several distinct relaxation rates ($\propto k_\text{eff}$) with different $D$ dependence.   This explains the limitations of existing scaling arguments based on a single-time scale~\cite{Austin04,Austin05,Reccius05}.  The predictions based on a scaling approach are reached only when $N$ is very large.  Thus simulations or experiments on short chains need to be interpreted with caution. Our approach can be extended to incorporate other non-trivial but important effects such as chain topology (e.g., branched polymers), electrostatics, molecular crowding, and chain stiffness~\cite{endnote_stiff}.          
Most importantly, the force-compression-extension relations presented above can be used to understand the elastic response of chromosomes confined in microfluidic devices, as well as the fate of proteins in nanopores.\\ 

We thank A. Arnold and D. Frenkel for helpful discussions in the early stage of this work and J. Bechhoefer for critical reading. The work of DT was supported by a grant from NSF CHE05-14056. BYH acknowledges financial  supports from the NSERC (Canada), and the BK 21 Frontier Physics Research Program during his stay at the Department of Physics and Astronomy, Seoul National University, Korea.

\appendix


\begin{thebibliography}{19}
\bibitem{endnote_Flory} 
The original Flory approach in $d$ dimensions leads to  $\beta \mathcal{F}\sim   \frac{R^2}{N a^2}+  \frac{a^d N^2}{R^d}$.  The renormalized Flory free energy in Eq.~\ref{eq:freeBY} can be obtained by setting $d=1$ and rescaling $a \rightarrow   D$ and $N \rightarrow  N/g$.  Linearity of $\mathcal{F}$ around $R =R_0$ in $d=1$ underlies the success of this rescaling scheme.
\bibitem{endnote_uncertainty}
In the experimental literature, even the basic de Gennes scaling of $R_0 \sim N D^{-\frac{2}{3}}$ has yet to be clarified for DNA in nanochannels, where significant fluctuations of $R_0$ (and, thus, the measurement uncertainties) are present~\cite{Austin05}.
\bibitem{endnote_dePablo}
Based on the results in Refs.~\cite{Thirum07, Arnold07b} and this study, 
care should be exercised in analyzing data on finite-sized DNA confined to pores.  The chain behavior can be sensitive to the precise choice of  the pair interaction of the chain beads especially if the channel diameter approaches the bead diameter.   This may be critical when coarse grained models are employed as in 
Ref.~\cite{dePablo03}.
\bibitem{endnote_I}
The emergence of the $D$-insensitive regime (green) is intriguing.  We believe that this can be attributed to the different setting for our theoretical analysis and/or partial chain backfolding.  Simulations on much longer chains are needed for further clarification.
\bibitem{endnote_stiff}
When $D$ is comparable to but larger than the persistence length, compression blobs can overlap each other relatively easily.  Chain stiffness can be implemented in our approach.
\end{thebibliography}
\end{document}